\documentclass[dvips]{article}
\usepackage{icrctc07}
\newcommand{\beq}{\begin{equation}}
\newcommand{\eeq}{\end{equation}}

\newcommand{\be}{\begin{equation}}
\newcommand{\ee}{\end{equation}}

\title{Evidence that a cluster of UHECRs was produced by a burst or flare}
\shorttitle{Evidence that a cluster...}

\authors{Glennys R. Farrar$^{1,2}$ }
\shortauthors{Farrar and et al.}
\afiliations{$^1$Center for Cosmology and Particle Physics, New York University\\ }
\email{gf25@nyu.edu}

\abstract{The angular clustering of 5 Ultrahigh Energy Cosmic Rays (UHECRs) in the combined published AGASA-HiRes data has a probability of $\sim 2 \times 10^{-3}$ of occurring by chance.  A first analysis of the implications of the event energies and angular spreading is presented, which is applicable if the source is close enough that GZK losses can be ignored.  Under this assumption, the observed energies of the events in this cluster favor a bursting rather than continuously emitting source, with the events emitted on a time scale short compared with $300 \, D_{\rm Mpc}$ years.  Assuming the UHECRs experience many incoherent small magnetic deflections enroute from source to Earth, the arrival direction distribution allows estimation that $\langle {B^2 \lambda} \rangle D \approx 7.7 {\rm \, nG^2 \, Mpc^2}$.  If the spectrum at the source $\sim E^{-2}$, the total isotropic equivalent energy emitted in UHECRs is $\ge 10^{43} \, D_{\rm Mpc}^3$ ergs.  
}

\begin{document}
\maketitle
\section{Introduction}

If UHECRs are charged particles and are produced by bursting sources, the events from a single burst should be clustered in energy as well as angle.  This is because high energy CRs experience lower deflection on average and therefore arrive to the observer more quickly on average than do lower energy events.  At any given place and time, the spectrum has about a factor of 3 spread in energies.  As time goes on, the average energy observed at any location decreases.  Here, I examine a cluster of 5 events in the combined published AGASA-HiRes data, whose spread in arrival directions is so small it could be consistent with instrumental resolution and is thus highly unlikely to be a chance clustering.  First, I compare the likelihood of fits assuming a continuous versus bursting source, then I assume the source is bursting and extract properties of the burst and intervening magnetic fields.  The analysis presented here ignores energy losses during propagation; a more comprehensive analysis is in preparation.

The complete published data from AGASA and the stereo HiRes consists of 57 AGASA events with nominal energy $\ge 40$ EeV and 271 HiRes events with nominal energy $\ge$ 10 EeV.  Both arrival directions and energies of the former are published\cite{AGASAupdate}; only arrival directions have been published for the latter\cite{HRclus04}.  With the HiRes collaboration, we searched for clusters of events in the combined high energy sample, 57 AGASA events above 40 EeV and 40 HiRes events above 30 EeV, using a Maximum Likelihood technique\cite{HRGF}.  One HiRes event at 38 EeV was found to be clustered with the known AGASA triplet, with an angular dispersion consistent with measurement resolution.  The probability of promoting the likelihood value of the triplet in the 57 event AGASA dataset, to that of the quadruplet in the combined 97 event dataset by chance, is $2 \times 10^{-3}$\cite{gfclus}.  This ``promotion probability" is a useful indicator of the significance since it is not skewed by the existence of the original triplet, in case that had been a chance occurrence.  Other measures of the likelihood of the quadruplet being a chance association give a similar significance\cite{gfclus}.  

A search of the remaining 231 HiRes events with energies in the range 10-30 EeV turns up a 5th event whose arrival direction is consistent with its coming from the same source with negligible magnetic deflection or dispersion.  Its association with the same source is less secure than for the original quadruplet, since there is a 1 in 6 chance of getting as high or higher likelihood value as observed in the data when adding 231 events at random to the 97 original events.   For the rest of the paper we proceed on the assumption that the 4 high energy events come from a single source, and will consider both cases that the 5th lower energy event is or is not from the same source.

\section{Energies favor a Bursting Source}
If UHECRs are produced by continuous sources and GZK distortions could be ignored, then the observed distribution of energies for a typical source would be the same as the distribution of all observed energies, and we could make a simple assessment of the probability that the cluster source is continuous, independent of knowledge of the exposure or the reliability of the energy measurements, since those affect the full dataset and the cluster equally on average.  Namely, the probability of selecting -- at random -- 4 events above 30 EeV and one or two events below, from a dataset with 97 events above 30 EeV and 231 events below.   If the cluster is taken to include only the 4 high energy events, the probability that their energies are drawn at random from the full data is $2 \times 10^{-4}$, while if all 5 events come from a single source the probability of finding the observed distribution of energies is $10^{-3}$.  Due to GZK distortions this analysis can however be quite misleading as will be discussed in a forthcoming paper.

In order to proceed in our analysis of the spectrum of an individual source, we need to know the total exposure to the cluster.  The  stereo HiRes exposure is energy dependent and has not been published, although the stereo exposure has been shown in conference talks and the mono exposures have been published.  The energy dependence of the stereo exposure should be very similar to that of a mono detector, so guided by figures shown at conferences, I normalize the integrated HiRes exposure, including weather cuts, to be equal at 70 EeV to the integrated AGASA exposure, ${\cal E}_{AG} = 1500 ~{\rm km^2 \, yr} $,  and I take the shape of the HiRes stereo exposure from the mono exposure using the functional form due to G. Hughes and D. Bergman (private communication), with the result
\beq 
{\cal E}_{HR}(E) = 7.5 \, 10^{-4} {\rm exp} [ c (1-e^{(-a (x-b))}) ] \, {\rm km^2 \, yr},
\eeq
where $x \equiv {\rm Log}_{10}(E /10^{16}{\rm eV}) $.  Above $E = 10^{17.8} {\rm eV}$ the parameters are $a$ =  1.175, $b$ = 0.732, and $c$ = 14.893.  While this is only an estimate of the HiRes stereo exposure, it is adequate for the purposes here.  The total exposure is ${\cal E} = {\cal E}_{HR}(E)+  {\cal E}_{AG} \,\theta(E-40 {\rm EeV})  $.  The relative exposure as a function of position, $\eta$,  is normalized such that $\int \eta \, d\Omega = 4 \pi$.  In the direction of the cluster it is about the same for both AGASA and HiRes: $\eta_c = 0.2 \, {\rm sr}^{-1}$.
 
In a beautiful paper in 1978, Alcock and Hatchett obtained the distribution in arrival time and direction of X-rays from an instantaneous point source and showed that it has a universal shape which depends on the source distance and parameters of the medium but not on details of the scattering process such as the form of the differential cross section\cite{alcockH}. Waxman and Miralda-Escude\cite{waxmanME} (WM-E) adapted the Alcock-Hatchett analysis to cosmic rays from a bursting source undergoing multiple small magnetic deflections.  WM-E derive the flux of UHECRs with energies in the range $\{ E, \, E + dE \}$ received from a burst which produced $N(E) dE$ cosmic rays in the same energy range, by an observer at a distance $D$ and time delay $\Delta t$ (relative to photons):
\be
\label{flux}
F(E; D, E_0) =  \frac{3 c E^2 N(E) }{8 \pi q^2 \langle B^2 \lambda \rangle D^4} G_{A\rm H}( (E/E_0)^2),
\ee
where $q$ is the charge, $\lambda$ is the characteristic length scale of the turbulent magnetic fields and $ \langle B^2 \lambda \rangle$ is defined precisely in \cite{waxmanME}.
An expression for $G_{\rm AH}$, the normalized probability distribution function, and also the joint distribution in energy and angle, are given in \cite{alcockH}.  Present statistics of this cluster are insufficient for the joint energy-angle distribution to be useful.   

\begin{figure}[t]
\begin{center}
\noindent
\includegraphics [width=0.5\textwidth]{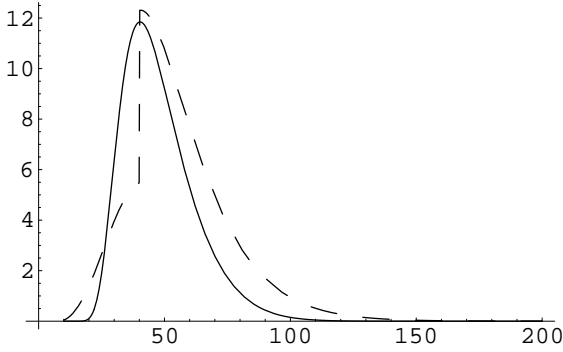}
\end{center}
\caption{
Observed spectrum for an $E^{-2}$ spectrum at the source and an energy-independent exposure (solid) or the actual exposure (dashed), for sources close enough that energy losses during propagation are insignificant.
}\label{fig:1}
\end{figure}

With an $E^{-2}$ spectrum at the source, the shape of the observed spectrum (\ref{flux}) is shown in Fig. \ref{fig:1} as a function of $E/E_0$, ignoring energy loss during propagation.   The parameter $E_0$ is determined by the distance and time delay of the observation, and the magnetic structure of the intervening medium:  
\be
\label{E0}
E_0 \equiv D  \left( \frac{2 q^2 \langle B^2 \lambda \rangle}{3 c \Delta t} \right)^{1/2}.
\ee
The peak of the spectrum is at $E_{\rm peak} =  0.214\, E_0$ and the average energy is $\bar{E} =\,  0.276 \,  E_0$.   

The characteristic spread in arrival directions averaged over all events of energy $E$ is 
\be
\label{thetas}
\theta_s(E) =  \frac{\sqrt {2 D q^2 \langle B^2 \lambda \rangle}}{3 E},
\ee
as for the case of a non-bursting source.  Keeping in mind that the time delay $\Delta t \sim D/c \, (1-cos(\theta_s)^2) \sim D \theta_s^2/(2 c)$ , the form of the expression (\ref{flux}) for the flux in each energy range can be easily understood:
\begin{eqnarray*}
\label{qualitative}
F_{\rm obs} &  \sim  \frac{{\rm (Number \,of \,events \,produced \,at \,the \,source})}{4 \pi D^2 \,{\rm (Spread \,in \,arrival \,times)}} \\
  & \sim \frac{N(E)}{4 \pi D^2 D \theta_s^2/(2 c)} \sim \frac{9 c q N(E) E^2}{4 \pi D^3 D  q^2 \langle B^2 \lambda \rangle },
\end{eqnarray*}
and the apparent difference in normalization compensates the peak magnitude of $G_{AH}$.

Note that when observing an instantaneous source at a {\it fixed time delay} -- as opposed to fixed energy -- the distribution of deflection angles with respect to the direction of the source is flatter than a gaussian for small angles and drops faster at large angles.  Waxman and Miralda-Escude\cite{waxmanME} obtained an approximate expression for the integral over energy of the joint AH distribution, at fixed time delay:
\be
\label{angdist}
P(\delta \theta \, ^2) \sim {\rm exp}(-0.58 \, [\,\delta \theta\,^2/\theta_s(\bar{E})^2]^2),  
\ee
where $\delta \theta$ is the angle between the arrival direction and the line of sight to the source.  We will use equation (\ref{angdist}) to fit for $\theta_s(\bar{E})$.  

Given the exposure to the source and a spectrum at the source, say $E^{-p}$, we can find the value of $E_0$ which gives the best fit to the data as follows.  Divide the energy range into $N$ bins.  For each bin the mean number of expected events is 
\be
\label{mui}
\mu_i(E_0) \equiv {\cal N}(E_0)\, E^{2-p}\, G_{\rm AH}((E/E_0)^2) \, {\cal E}(E),
\ee where the normalization factor ${\cal N}(E_0)$ is chosen so that $\Sigma_{i = 1}^N \mu_i(E_0) \equiv N_c$ is the number of events in the cluster.  The most probable value of $E_0$ is the one which maximizes the likelihood measure 
\be
\label{LM}
LM \equiv \Pi_{i = 1}^N {\cal P}(\mu_i(E_0),n_i),
\ee
where $ {\cal P}(\mu_i,n_i)$ is the Poisson probability of finding the observed number $n_i$ of events in the $i$th bin when $\mu_i$ are expected.  We adopt henceforth
$p = 2$  for definiteness.  This choice simplifies formulae and is likely to be close to reality, and the inferred $E_0$ differs only slightly with other choices, e.g., $p=2.7$.  

The energy of the 5th, low energy event has not been published, but HiRes has kindly released the event's ranking in energy (150th), constraining its energy to be within a few EeV of 15 EeV; this is plenty accurate for our needs.  Using just the four highest energy events, the best fit value is $E_0^{(4)} = 190$ EeV; with all five events $E_0^{(5)} = 150$ EeV.   

The actual value of the likelihood measure defined above has no particular significance -- for instance, it depends on the number of bins -- but it can be used to assess the relative quality of different fits to the spectrum of the cluster by taking the ratio of the $LM$ values.  GZK distortions must be included for this to be meaningful for all source distances; the results will be presented elsewhere.


\section{Angular Distribution and Flux}

Henceforth we assume the events in the cluster are all protons.  The best fit to the observed arrival directions under the bursting source hypothesis is obtained by maximizing the product of the angular probability densities for the $N_c$ events of the cluster, with respect to the direction of the source $\vec{\theta_0}$ and the energy-dependent magnetic smearing parameter $\theta_s(E)$ introduced in (\ref{thetas}).  For each event, the 2-d probability density is
proportional to
\beq
{\rm exp} \left[-0.58 \left( \frac{ ( \vec{\theta_i} - \vec{\theta_0})^2}{ \theta_s(\bar{E})^2} \right)^2 \right] {\rm exp}\left[-  \frac{ (\vec{\theta_i} - \vec{\theta_0})^2}{ \sigma_i^2}\right].
\eeq
This results in a source direction \{RA, dec\} $= \{169.67^\circ, \,56.70^\circ \}$ and 
\be
\label{thetas}
\theta_s(\bar{E}) = 1.14^\circ.
\ee
From these values one finds
\be
\label{BsqlamD}
\langle {B^2 \lambda} \rangle D_0 \approx 7.0\, {\rm EeV^2} = 7.7\, {\rm nG^2 \,Mpc^2},
\ee
where $D_0$ is the distance to the source which produced the quad.

The normalization of the spectrum at the source can be estimated in units of $ \langle B^2 \lambda \rangle D^4$ either by equating the total number of events, $N_c$, or energy received from the source, $\Sigma_{i=1}^{N_c} E_i$, to $\int E^{0,1} F(E) {\cal E}(E) dE$, respectively.  This gives
\be
\label{k}
\frac{N(E) E^2 3 c}{8 \pi \langle B^2 \lambda \rangle D^4} = 2.4 \,(...) \, 10^{-5}
\ee
respectively.  Using the value of $\langle B^2 \lambda \rangle D_0$ from the angular distribution gives
\be
\label{norm}
E^2 N(E) = 4.0 \, 10^{42} D_{\rm Mpc}^3 \, \left( \frac{\theta_s}{1.14^\circ } \right)^2 { \rm erg}.
\ee 
Integrating this over energy yields the isotropic equivalent energy the source emitted in ultrahigh energy cosmic rays with energies in the range 10-300 EeV: 
\be
\label{EUHECR}
E_{10-300} = 1.3 \, 10^{43}\left( \frac{\theta_s}{1.14^\circ} \right)^2 \, D_{Mpc}^3 \,  { \rm erg }.
\ee
If the UHECRs from the source are beamed toward us in a cone of solid angle $\Delta \Omega$, the total energy in UHECRs should be reduced by the factor $  \frac{\Delta \Omega}{4 \pi} $.  

Since GZK losses have not been included, (\ref{EUHECR}) can be considered a lower limit to the energy in UHECRs at the source.  


\section{Minimum Flare Duration}
From eqns \ref{E0} and \ref{thetas} and the fit to $\theta_s$ we have that the arrival time delay is 
\be
\label{tau}
\Delta t = \frac{2}{9} \theta_s^2 \frac{D_0}{c} \approx 300 \, D_{\rm Mpc} \,{\rm  yr}.
\ee
The source can be flaring rather than bursting and lead to the observed spectrum, as long as the source's peak luminosity is much greater than its normal or quiescent luminosity, for a timescale short compared to $\Delta t$.

\section{Conclusion}

The analysis presented here, ignoring GZK distortions, provides a first, naive look at the implications of the Ursa Major event energies and angular separations for the nature of the source.  It serves as a guide to the more complex analysis required when the source is at a large enough distance that GZK losses are important; the analysis and simulations for the more general case will be presented elsewhere.  If the source is close enough for this analysis to be applicable, the spectrum favors a bursting or flaring rather than continuous source.  In that case, the total isotropic equivalent energy emitted in CRs in the range $10^{19}-3 10^{20}$ eV is $\ge 10^{43} \, D_{\rm Mpc}^3$ ergs and the duration of the flare is $<< 300 \, D_{\rm Mpc}$ years.

\bibliography{icrc1269}
\bibliographystyle{plain}
\end{document}